\documentclass[pre,aps,reprint,twocolumn,showpacs,floatfix]{revtex4-1}

\usepackage[dvips]{graphicx}
\usepackage{amsmath}
\usepackage{amssymb}
\usepackage[nice]{nicefrac}
\usepackage{color}

\newcommand{\MFPTb}{T^{\mathrm{w-b}}_{\mathrm{MFPT}}}
\newcommand{\MFPTw}{T^{\mathrm{w-w}}_{\mathrm{MFPT}}}
\newcommand{\MFPT}{T_{\mathrm{MFPT}}}

\newcommand{\Sb}{S^{\mathrm{w-b}}(t)}
\newcommand{\Sw}{S^{\mathrm{w-w}}(t)}

\newcommand{\fb}{f^{\mathrm{w-b}}(t)}
\newcommand{\fw}{f^{\mathrm{w-w}}(t)}

\newcommand{\xw}{x_{\mathrm{w}}}

\newcommand{\xb}{x_{\mathrm{b}}}

\begin{document}

\title{Escape from the potential well: competition between long jumps and long waiting times}

\author{Bart{\l}omiej Dybiec}
\email{bartek@th.if.uj.edu.pl}
\affiliation{Marian Smoluchowski Institute of Physics, and Mark Kac Center for Complex Systems Research, Jagellonian University, ul. Reymonta 4, 30--059 Krak\'ow, Poland}

\date{\today}
\begin{abstract}
Within a concept of the fractional diffusion equation and subordination, the paper examines the influence of a competition between long waiting times and long jumps on the escape from the potential well. Applying analytical arguments and  numerical methods, we demonstrate that the presence of long waiting times distributed according to a power-law distribution with a diverging mean leads to very general asymptotic properties of the survival probability. The observed survival probability asymptotically decays like a power-law whose form is not affected by the value of the exponent characterizing the power-law jump length distribution. It is demonstrated that this behavior is typical of and generic for systems exhibiting long waiting times. We also show that the survival probability has a universal character not only asymptotically but also at small times. Finally, it is indicated which properties of the first passage time density are sensitive to the exact value of the exponent characterizing the jump length distribution.
\end{abstract}

\maketitle

%
%
\section{Introduction}

Kramers' seminal paper \cite{kramers1940} studies the problem of the escape of a particle over the potential barrier. An escape of the test particle from the metastable state takes place due to the presence of noise. Otherwise, in the static potential, any particle (described by the Newtonian mechanics) would be sliding to the nearest minimum of the potential only.  Random interactions of the test particle with other particles of the medium make the escape from the metastable state possible.

In the simplest situations, it is assumed that the noise acting on a test particle is white and Gaussian.
In far from equilibrium situations, the stochastic process describing interactions of the test particle with the environment can be still of the white type but its values can be distributed according to some heavy tailed distribution. An $\alpha$-stable noise, leading to L\'evy flights \cite{janicki1994,*nolan2010,chechkin2006,dubkov2008}, can be considered \cite{chechkin2005,dybiec2007,Chechkin2007} as an exemplary driving process.
L\'evy stable noise is especially suitable for description of the far from equilibrium situations because of its heavy-tail characteristics, infinite divisibility and the presence of the generalized central limit theorem \cite{janicki1994,*nolan2010}. The presence of $\alpha$-stable L\'evy type noises affects the escape process. However, the survival probability, as in the case of the white Gaussian noise, still has the exponential asymptotic form \cite{eliazar2004,chechkin2005,imkeller2006b,dybiec2007,Chechkin2007}. L\'evy flights which have been observed in various real-life situations when a test particle is able to perform unusually large jumps \cite{dubkov2008,shlesinger1995,metzler2000,*metzler2004}. L\'evy flights have been documented to describe motion of fluorescent probes in living polymers, tracer particles in rotating flows \cite{solomon1993,*solomon1994} and cooled atoms in laser fields. They also serve as a paradigm of efficient searching strategies on folding polymer chains \cite{sokolov1997,*lomholt2005,*belik2007} and environmental problems \cite{Edwards2007,*reynolds2009}.

The deviations from the exponential asymptotic form of the survival probability can be introduced by the trapping events distributed according to a power-law distribution with the diverging mean \cite{metzler2000,Metzler2000c}, see Sec.~\ref{sec:results}. Transport in porous, fractal-like media or relaxation kinetics in inhomogeneous materials lead to long trapping events which can be distributed according to a power-law. Consequently, such processes are usually ultraslow, i.e. subdiffusive \cite{metzler2000,*metzler2004}. The most intriguing situation takes place, however, when both effects -- the presence of long waiting times for the next step and long jumps -- are combined in the same scenario, see \cite{zumofen1995,metzler2000,klages2008,dybiec2009h,ai2010}, because long jumps can hide  temporal dynamics induced by long waiting times. This competition can be of special relevance in single particle tracking experiments \cite{lubelski2008,*he2008,*metzler2009,*bronstein2009}. For example, a particle moving along a polymer can be trapped in a series of local minima of the potential while long jumps can take place in conformational space of the fast folding polymer \cite{brockmann2002,lomholt2005,*belik2007} where the distance measured along the polymer can be significantly longer than the Euclidean distance. In such a case instead of travelling a long distance along the polymer,  due to chain looping and volume exchange \cite{lomholt2005}, it is possible to jump directly to the other segment of the polymer.

In the region of non-Markovian L\'evy flights \cite{metzler2000}, i.e. continuous time random walks with power-law waiting time $p(\Delta t)\propto t^{-(\nu+1)}$ ($0<\nu<1$) and jump length $p(\Delta x)\propto |x|^{-(\alpha+1)}$ ($0<\alpha<2$) distributions , the evolution of the probability density of finding a random walker at the time $t$ in the vicinity of $x$ is asymptotically described by the fractional diffusion equation \cite{metzler1999,metzler2000,*metzler2004}
\begin{equation}
 \frac{\partial p(x,t)}{\partial t}={}_{0}D^{1-\nu}_{t}\left[ \frac{\partial}{\partial x} V'(x) + \sigma^\alpha \frac{\partial^\alpha}{\partial |x|^\alpha} \right] p(x,t),
\label{eq:ffpe}
\end{equation}
where $V(x)$ represents external, static potential. In Eq.~(\ref{eq:ffpe}), $\partial^\alpha/\partial |x|^\alpha$ stands for the Riesz fractional (space) derivative which is defined via the Fourier transform
${\cal{F}}\left[\frac{\partial^{\alpha} f(x)}{\partial |x|^{\alpha}} \right]=-|k|^{\alpha}\mathcal{F}\left[{f}(x)\right],$
while ${}_{0}D^{1-\nu}_{t}$ denotes the Riemann-Liouville fractional (time) derivative ${}_{0}D^{1-\nu}_{t}=\frac{d}{d t}{}_{0}D^{-\nu}_{t}$ defined as
${}_{0}D^{1-\nu}_{t}f(x,t)=\frac{1}{\Gamma(\nu)}\frac{d}{d t}\int^{t}_0 dt'\frac{f(x,t')}{(t-t')^{1-\nu}}$
\cite{podlubny1998,metzler2000}.

Here, we focus on the investigation of the influence of a competition between long waiting times and long jumps on the Kramers problem \cite{kramers1940,pollak1986,hanggi1990,banik2000}. As a jump length distribution we chose symmetric $\alpha$-stable densities which are responsible for the occurrence of the fractional space derivative in Eq.~(\ref{eq:ffpe}) \cite{fogedby1994,*fogedby1994b,*compte1996,yanovsky2000,*schertzer2001}. Symmetric $\alpha$-stable densities asymptotically behave like a power-law, see \cite{janicki1994,*nolan2010,chechkin2006,dubkov2008}. Nevertheless, what is even more important is the fact that for jump  lengths $p(\Delta x)$ distributed according to symmetric $\alpha$-stable densities the CTRW scenario leads exactly to Eq.~(\ref{eq:ffpe}). The parameter $\alpha$ ($0 < \alpha \leqslant 2$) is called the stability index, see \cite{janicki1994,*nolan2010}. In the limit of $\alpha = 2$ any $\alpha$-stable density is equivalent to the Gaussian distribution. Therefore, in the limit of $\alpha=2$ the fractional (space) derivative in Eq.~(\ref{eq:ffpe}) is replaced by the partial derivative of the second order. The parameter $\nu$ ($0<\nu < 1$), see Eq.~(\ref{eq:ffpe}), is responsible for the trapping events and is traditionally called the subdiffusion parameter \cite{metzler2000}. In the limit of $\nu = 1$ the process described by Eq.~(\ref{eq:ffpe}) becomes a Markovian diffusion process and the fractional (time) derivative disappears. Breaking of Markovianity is introduced by the long waiting times which are described by the fractional (time) derivative \cite{podlubny1998,metzler2000}. Finally, in the limit of $\alpha>2$ and $\nu>1$ the mean waiting time and variance of the jump length becomes finite and the fractional Fokker-Planck equation is replaced by the standard Fokker-Planck equation \cite{shlesinger1982,risken1984,klafter1987,gardiner2009}.

\begin{figure}[!ht]
\includegraphics[angle=0, width=8.0cm, height=5cm]{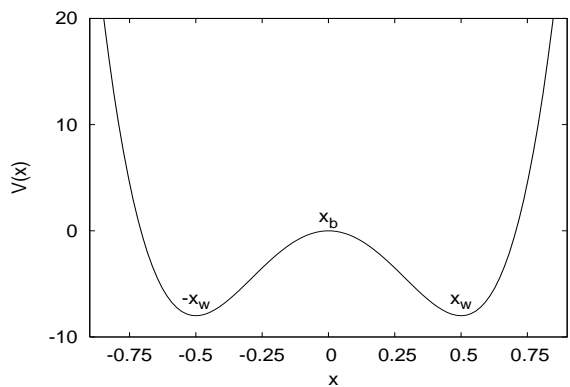}
\caption{The Kramers problem is studied for the generic double-well, fourth-order potential
$V(x)=-\frac{a}{2}x^2+\frac{b}{4}x^4$. Initially, a test particle is located in the left potential well.}
\label{fig:potential}
\end{figure}

%
%
\section{Methods and results\label{sec:results}}

We extend earlier studies \cite{ditlevsen1999,chechkin2005,dybiec2007} by investigating the influence of the competition between long waiting times and long jumps on the Kramers problem \cite{kramers1940,pollak1986,hanggi1990,banik2000}. Initially, at the time $t=0$ a test particle is located in the left potential well of the generic, fourth-order, double-well potential
\begin{equation}
V(x)=-\frac{a}{2}x^2+\frac{b}{4}x^4.
\label{eq:potential}
\end{equation}
The minima of the potential, see Eq.~(\ref{eq:potential}) and Fig.~\ref{fig:potential}, are located at $\pm \xw= \pm \sqrt{a/b}$. The height of the potential barrier is $\Delta V = V(0)-V(\xw)=a^4/4b$. We study and compare two different setups: (i) escape from the potential well to the barrier top ($\xb$) and (ii) escape from the potential well to the (other) potential well ($\xw$)  \cite{dybiec2007}, i.e. we assume that the border between the states is located either at the barrier top or at the other potential well, see Fig.~\ref{fig:potential}.

In the case of (Markovian) L\'evy flights ($\nu=1$ with $\alpha<2$) Eq.~(\ref{eq:ffpe}) reduces to
\begin{equation}
 \frac{\partial p(x,t)}{\partial t}=\left[ \frac{\partial}{\partial x} V'(x) + \sigma^\alpha \frac{\partial^\alpha}{\partial |x|^\alpha} \right] p(x,t).
\label{eq:ffpelf}
\end{equation}
The stochastic representation of solutions of Eq.~(\ref{eq:ffpelf}) is provided by the following Langevin equation
\begin{equation}
dX(t)=-V'(X(t))dt+ dL_{\alpha,0}(t),
\label{eq:langevineqlf}
\end{equation}
driven by the symmetric $\alpha$-stable motion $L_{\alpha,0}(t)$ whose increments, $\Delta L_{\alpha,0}(\Delta t)=L_{\alpha,0}(t+\Delta t)-L_{\alpha,0}(t)$, are independent and distributed according to the symmetric $\alpha$-stable density with the corresponding characteristic function $\phi(k) = \langle  \exp(ik \Delta L_{\alpha,0}) \rangle = \exp\left( -\Delta t\sigma^\alpha|k|^\alpha \right)$ \cite{janicki1994,*nolan2010,chechkin2006,dubkov2008}. Here the parameter $\alpha\in(0,2]$ denotes the stability index, describing the asymptotic long-tail power-law for the $x$-distribution, which for $\alpha<2$ is of the $|x|^{-(1+\alpha)}$ type. The parameter $\sigma$ ($\sigma\in(0,\infty)$) is the scale parameter which characterizes the overall distribution width.

For the exponent characterizing the waiting time distribution $\nu<1$, the stochastic representation of solutions of Eq.~(\ref{eq:ffpe}) is provided by the subordination method  \cite{magdziarz2007b,magdziarz2007}, i.e.
the process inspected $X(t)$ is obtained as a function $X(t)=\tilde{X}(S_t)$ by using a different (stochastic) clock $S_t$. In this framework $S_t$ denotes the $\nu$-stable subordinator,
$S_t=\mathrm{inf}\left \{s:U(s)>t \right \}$, where $U(s)$ stands for a strictly increasing $\nu$-stable process whose distribution has the Laplace transform $\langle e^{-kU(s)} \rangle =e^{-sk^{\nu}}$. The parent process $\tilde{X}(s)$ is composed of independent increments of the symmetric $\alpha$-stable motion described in an operational time~$s$
\begin{equation}
d\tilde{X}(s)=-V'(\tilde{X}(s))ds+dL_{\alpha,0}(s).
\label{eq:langevineq}
\end{equation}
The subordination provides a proper stochastic realization of the random process described by the fractional Fokker-Planck equation (\ref{eq:ffpe}) \cite{magdziarz2007b,magdziarz2007}. The stochastic representation of solutions of Eq.~(\ref{eq:ffpe}), see \cite{magdziarz2007b,magdziarz2007}, is valid not only asymptotically (large $x$ and long $t$) but for all times $t$ and positions $x$ \cite{magdziarz2007b,magdziarz2007}. The stochastic representation of solutions of Eq.~(\ref{eq:ffpe}) based on the subordination method does not use explicitly the waiting time distribution. The information about waiting time distribution is implicitly incorporated in the $\nu$-stable subordinator $S_t$ which provides a link between the physical time $t$ and the operational time $s$ \cite{magdziarz2007b,magdziarz2007}. However, there are other, asymptotic, stochastic representation of solutions of Eq.~(\ref{eq:ffpe}) which make explicit use of the waiting time distribution \cite{heinsalu2006,fulger2008}, when indeed the waiting time distribution has the power-law asymptotics.

The first passage time density is related to the probability density $p(x,t)$, see Eq.~(\ref{eq:ffpe})
\begin{equation}
f^{\mathrm{w-x}}(t)=-\frac{d}{dt}\int_{-\infty}^{x} p(x,t)dx=-\frac{d}{dt} S^{\mathrm{w-x}}(t),
\label{eq:fptd}
\end{equation}
where $S^{\mathrm{w-x}}(t)=\int_{-\infty}^{x} p(x,t)dx$ is the survival probability, i.e. the probability that at the time $t$ a test particle, which started its motion at the left potential well, has not been yet absorbed at the boundary located at the barrier top $\xb$ or at the potential well $\xw$. In more general cases, when trajectories become discontinuous, boundary conditions are non-local, see below. The mean first passage time can be calculated as
\begin{equation}
\MFPT^{\mathrm{w-x}}=\int_0^\infty t f^{\mathrm{w-x}}(t) dt = \int_0^\infty S^{\mathrm{w-x}}(t)dt,
\end{equation}
where $x$ stands for the final point, i.e. the barrier top ($\xb$) or the potential well ($\xw$). In order to guarantee the finite value of the mean first passage time, the survival probability (as well as the first passage time density) asymptotically has to tend to zero fast enough. In particular, for $\nu=1$ and $\alpha=2$, the standard Kramers problem \cite{kramers1940} is recovered, the survival probability has the exponential asymptotics and the mean first passage time can be calculated exactly \cite{gardiner2009}. The survival probability attains the exponential asymptotics \cite{eliazar2004,chechkin2005,imkeller2006b,dybiec2007,Chechkin2007} also for Markovian L\'evy flights when the mean first passage time is finite \cite{dybiec2007,Chechkin2007}. However, due to memory effects for $\nu<1$  the asymptotic behavior of the survival probability is of the power-law type, see below. In such a case, in order to assure a finite value of the mean first passage time, the exponent characterizing the decay of the survival probability needs to be larger than one, i.e. $S^{\mathrm{w-x}}(t) \propto t^{-\gamma}$ with $\gamma > 1$. By analogy, the decay of the first passage time density needs to be characterized by the exponent larger than two. Otherwise, the mean first passage time would be infinite due to divergence at infinity.

Properties of the first passage time distribution can be determined from Eqs.~(\ref{eq:ffpe}) and (\ref{eq:fptd}). The solution $p(x,t)$ of Eq.~(\ref{eq:ffpe}) can be written \cite{risken1984,gardiner2009,metzler2000} as a sum of eigenfunctions
\begin{equation}
p(x,t)=\sum_{i}c_ip_i(x,t)=\sum_{i}c_iT_i(t)\varphi_i(x)
\label{eq:factorization}
\end{equation}
where $T_i(t)$ fulfills
\begin{equation}
 \frac{dT_i(t)}{dt}=-\lambda_{i,\nu}\; {}_{0}D^{1-\nu}_{t} T_i(t)
\label{eq:timeeigenfunction}
\end{equation}
while $\varphi_i(x)$ fulfills
\begin{equation}
\left[ \frac{\partial}{\partial x} V'(x) + \sigma^\alpha \frac{\partial^\alpha}{\partial |x|^\alpha} \right]\varphi_i(x) = -\lambda_{i,\nu} \varphi_i(x).
\label{eq:spaceeigenfunction}
\end{equation}
The solution of Eq.~(\ref{eq:timeeigenfunction}) is given by the Mittag-Leffler function \cite{podlubny1998}
\begin{equation}
 T_i(t)=E_\nu(-\lambda_{i,\nu} t^\nu) \equiv \sum_{j=0}^\infty \frac{(-\lambda_{i,\nu} t^\nu)^j}{\Gamma(1+\nu j)}.
\end{equation}
Eigenvalues $\lambda_{i,\nu}$ are defined by boundary conditions to Eq.~(\ref{eq:spaceeigenfunction}) and coefficients $c_i$ are determined by the initial condition. More precisely, at the barrier top ($\xb$) or at the potential well ($\xw$) there are absorbing boundaries. Consequently, for $\alpha=2$ one has $p(x = \xb,t)=0$ or $p(x = \xw,t)=0$ while for $\alpha<2$ (due to discontinuity of trajectories) one has $p(x \geqslant \xb,t)=0$ or $p(x \geqslant \xw,t)=0$, see \cite{dybiec2006,*zoia2007}.

The main features of the first passage time distribution can be deducted without the knowledge of the eigenvalues $\lambda_{i,\nu}$ because of the properties of the Mittag-Leffler function. The Mittag-Leffler function has a stretched exponential form for small values of the argument
\begin{equation}
 E_\nu(-t^\nu) \propto \exp\left[ -\frac{t^\nu}{\Gamma(1+\nu)} \right]
\label{eq:mlfstretched}
\end{equation}
while asymptotically for large arguments it behaves like
\begin{equation}
 E_\nu(-t^\nu) \propto t^{-\nu},
\label{eq:mlfasymptotics}
\end{equation}
see \cite{metzler2000}.
For $\nu<1$, the asymptotic behavior of the Mittag-Leffler function is solely determined by the value of the exponent $\nu$. In the limit of $\nu=1$ the Mittag-Leffler function is equivalent to the exponential function.

The survival probability, $S(t)$, and the first passage time density, $f(t)$, can be calculated by use of Eqs.~(\ref{eq:fptd}) and (\ref{eq:factorization}). The integration of the spatial eigenfunctions $\varphi_i(x)$ can be considered as a modification of constants $c_i$, see Eq.~(\ref{eq:factorization}). Therefore, $S(t)$ is a sum (with unknown coefficients) of the temporal eigenfunctions $T_i(t)$. For $\nu=1$, the process studied is Markovian and the survival probability has the asymptotic exponential behavior which is determined by the (unknown) smallest eigenvalue $\lambda_\mathrm{min}=\min_{i}\{ \lambda_i  \}$
\begin{equation}
 S_{\nu=1}(t) \propto \exp(-\lambda_\mathrm{min} t).
\label{eq:sasymptotic1}
\end{equation}
For $\nu<1$, the process studied is non-Markovian and  the survival probability has the asymptotic power-law behavior (see Eq.~(\ref{eq:mlfasymptotics})) determined by the exponent $\nu$
\begin{equation}
 S(t) \propto t^{-\nu}.
\label{eq:sasymptotic}
\end{equation}

\begin{figure}[!ht]
\begin{center}
\includegraphics[angle=0,width=0.99\columnwidth]{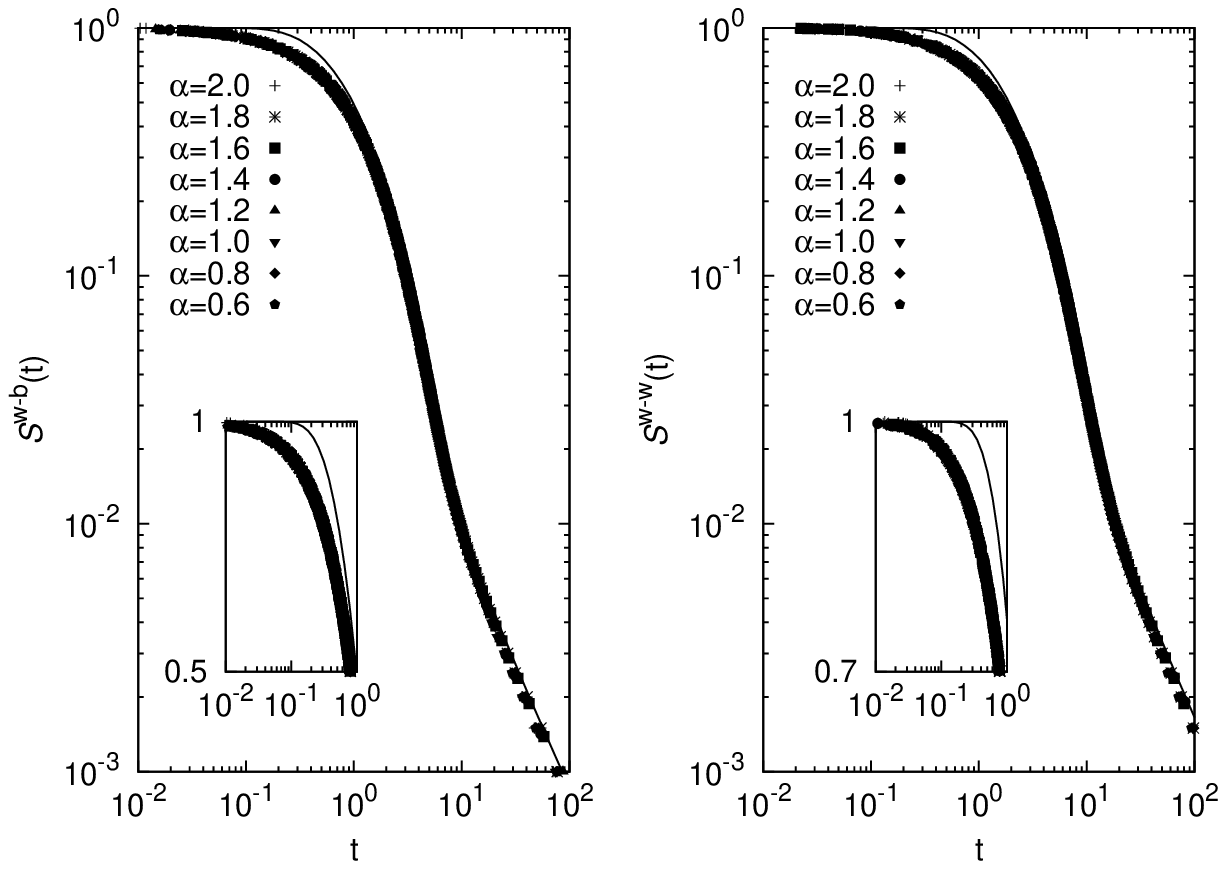}
\caption{The rescaled survival probability $S(t)$ for the Kramers problem in the case of non-Markovian L\'evy flights. The absorbing boundary is located at the barrier top $\xb$ ($\Sb$ -- left panel) or at the right potential well $\xw$ ($\Sw$ -- right panel). The value of the waiting time exponent is $\nu=0.95$. Various symbols correspond to various values of the stability index $\alpha$. The solid lines present the survival probability for the escape from the finite interval of the length $L=1$, see \cite{dybiec2010}.}
\label{fig:fptd095}
\end{center}
\end{figure}

\begin{figure}[!ht]
\begin{center}
\includegraphics[angle=0,width=0.99\columnwidth]{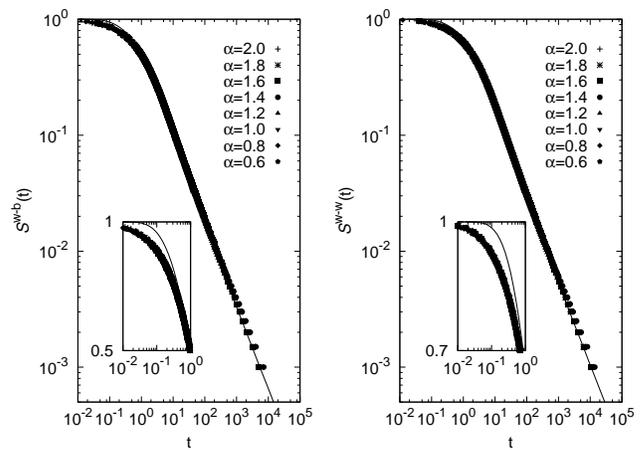}
\caption{The same as in Fig.~\ref{fig:fptd095} for the waiting time exponent $\nu=0.7$. The asymptotic power-law behavior, characterized by exponent $-\nu$, is well visible.}
\label{fig:fptd07}
\end{center}
\end{figure}

\begin{figure}[!ht]
\begin{center}
\includegraphics[angle=0,width=0.99\columnwidth]{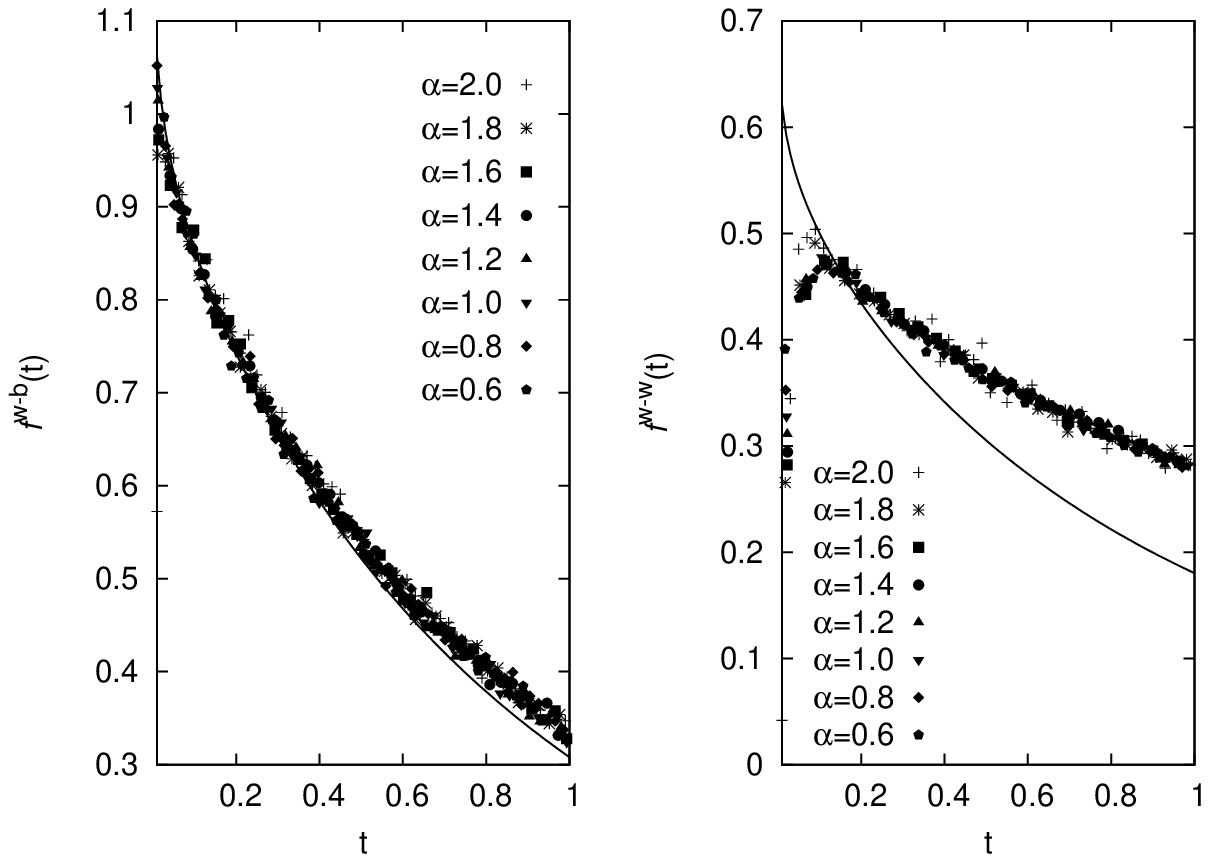}
\caption{
The rescaled first passage time density $f(t)$ for the Kramers problem in the case of non-Markovian L\'evy flights. The absorbing boundary is located at the barrier top $\xb$ ($\fb$ -- left panel) or at the right potential well $\xw$ ($\fw$ -- right panel). The value of the waiting time exponent is $\nu=0.95$.  Various symbols correspond to various values of the stability index $\alpha$. The solid lines present short time behavior of the derivative of the Mittag-Leffler function.}
\label{fig:his095res}
\end{center}
\end{figure}

\begin{figure}[!ht]
\begin{center}
\includegraphics[angle=0,width=0.99\columnwidth]{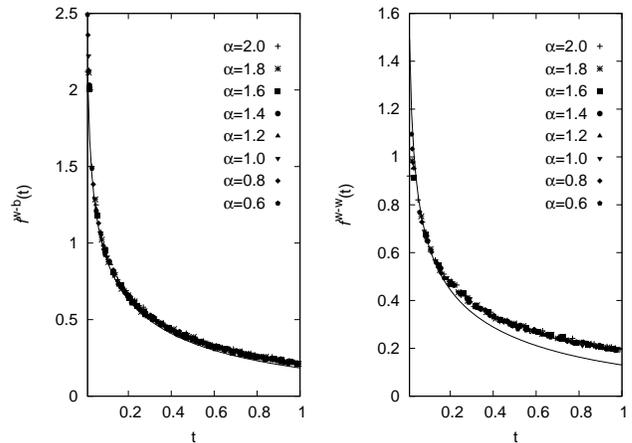}
\caption{The same as in Fig.~\ref{fig:his095res} for the waiting time exponent $\nu=0.7$.}
\label{fig:his07res}
\end{center}
\end{figure}

\begin{figure}[!ht]
\begin{center}
\includegraphics[angle=0,width=0.99\columnwidth]{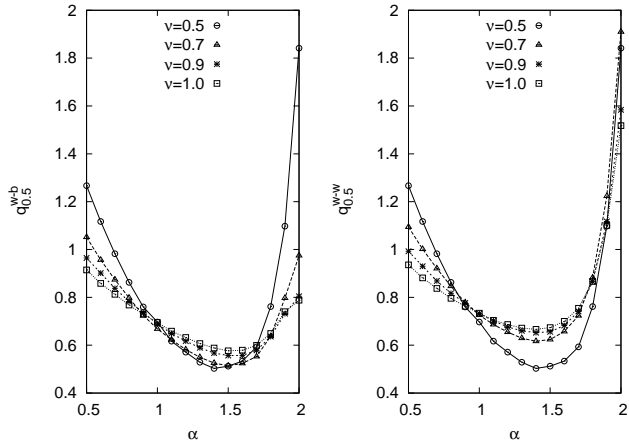}
\caption{Locations of the medians (quantile $q_{0.5}$) of the first passage time densities as the function of the stability index $\alpha$ for the potential well to the barrier top ($q^{\mathrm{w-b}}_{0.5}$ -- left panel) and the potential well to the potential well ($q^{\mathrm{w-w}}_{0.5}$ -- right panel) setups. Various symbols correspond to various values of the waiting time exponent $\nu$. Lines are drawn to guide the eye only.}
\label{fig:medianaalpha}
\end{center}
\end{figure}

\begin{figure}[!ht]
\begin{center}
\includegraphics[angle=0,width=0.99\columnwidth]{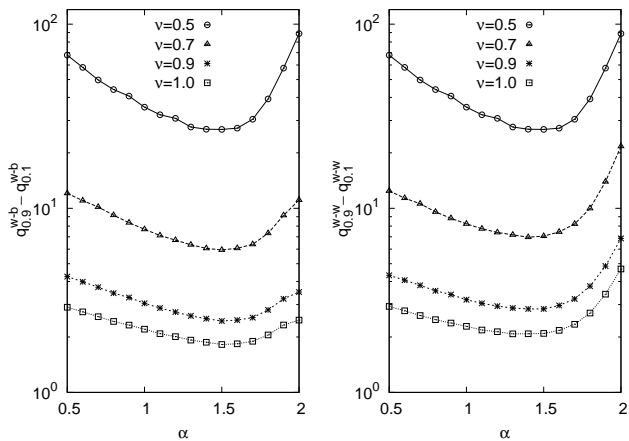}
\caption{The width of the first passage time density defined as the interquantile distance ($q_{0.9}-q_{0.1}$) as the function of the stability index $\alpha$ for the potential well to the barrier top ($q^{\mathrm{w-b}}_{0.9}-q^{\mathrm{w-b}}_{0.1}$ -- left panel) and the potential well to the potential well ($q^{\mathrm{w-w}}_{0.9}-q^{\mathrm{w-w}}_{0.1}$ -- right panel) setups. Various symbols correspond to various values of the waiting time exponent $\nu$. Lines are drawn to guide the eye only.}
\label{fig:widthalpha}
\end{center}
\end{figure}

\begin{figure}[!ht]
\begin{center}
\includegraphics[angle=0,width=0.99\columnwidth]{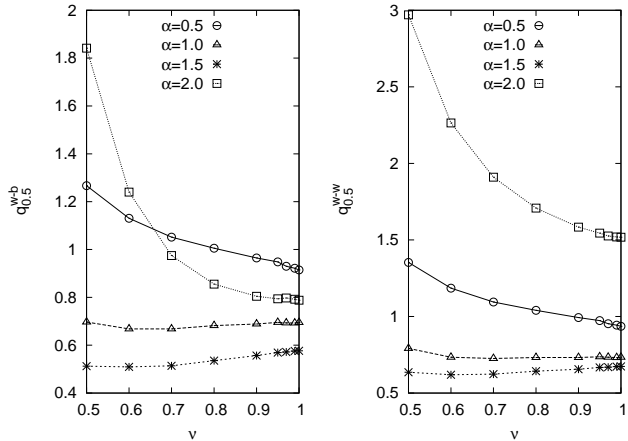}
\caption{The same as in Fig.~\ref{fig:medianaalpha} as the function of the waiting time exponent $\nu$. Various symbols correspond to various values of the stability index $\alpha$. Lines are drawn to guide the eye only.}
\label{fig:mediananu}
\end{center}
\end{figure}

\begin{figure}[!ht]
\begin{center}
\includegraphics[angle=0,width=0.99\columnwidth]{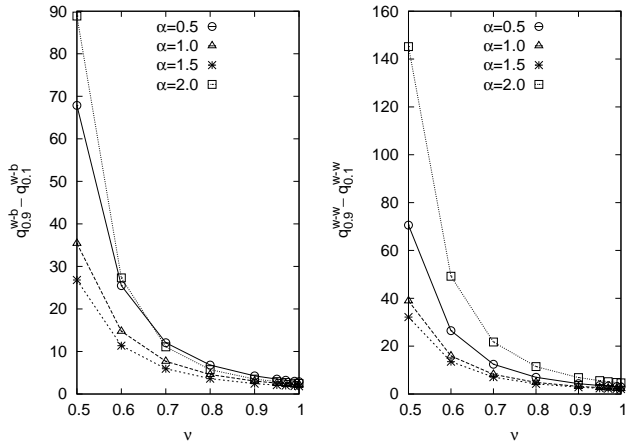}
\caption{The same as in Fig.~\ref{fig:widthalpha} as the function of the waiting time exponent $\nu$. Various symbols correspond to various values of the stability index $\alpha$. Lines are drawn to guide the eye only.}
\label{fig:widthnu}
\end{center}
\end{figure}

Figures~\ref{fig:fptd095}--\ref{fig:widthnu} present numerical results obtained using the subordinated Langevin equation, see Eq.~(\ref{eq:langevineq}) and \cite{magdziarz2007b,magdziarz2007}. Numerical simulations were performed with the same set of parameters as in \cite{dybiec2007}, i.e. $\sigma=\sqrt{2}$, $a=128$ and $b=512$ (see Eq.~(\ref{eq:potential}) and Fig.~\ref{fig:potential})). Such a choice of parameters leads to $\xw= 1/2$ and $\Delta V=8$, i.e. the potential wells are located at $\pm 1/2$ while the barrier height is $8$. Numerical simulations have been performed with the time step of integration $\Delta t=10^{-4}$ and averaged over $N=3 \times 10^5$ realizations. The range of parameters has been restricted to $0.5 \leqslant \alpha \leqslant 2$  and $ 0.5 \leqslant \nu \leqslant 1$, because a small value of the stability index $\alpha$ might lead to numerical inaccuracy while small values of the waiting time exponent result in a significant increase of the simulation time. In the limit of $\nu=1$, numerical methods applied perfectly recovered all the results of earlier studies \cite{dybiec2007}. Therefore, the Monte Carlo inspection of the Markovian Kramers problem served as a benchmark for implemented numerical methods.

In order to estimate the survival probability it is necessary to collect the sample of first passage times. Required first passage times are obtained by numerical simulation of the subordinated Langevin equation \cite{janicki1994,janicki1996,dybiec2004b,dybiec2006,magdziarz2007b,magdziarz2007}. The Langevin equation (\ref{eq:langevineq}) is simulated until the first escape from the domain of motion, i.e. as long as $x(t)<\xb$ or $x(t)<\xw$. For $\alpha<2$, trajectories of the process $x(t)$ are discontinuous. Consequently, boundary conditions become non-local \cite{dybiec2006,*zoia2007} because the barrier location is not hit by a majority of trajectories. Therefore, in order to assure correct evaluation of the first passage time it is necessary to impose non-local boundary conditions. Within simulations it is assumed that the entire semi line $x \geqslant \xb$ or $x \geqslant \xw$ is absorbing. The discontinuity of trajectories is responsible for failure of the method of images \cite{chechkin2003b,*sokolov2004b,metzler2000,*metzler2004} and the presence of leapovers \cite{koren2007,*koren2007b}.  Due to leapovers, the particle escaping from the initial potential well can jump to a distant point located far beyond the position of the absorbing boundary. With the non-negligible probability, the jump can be long enough to overpass simultaneously both locations of absorbing boundaries $\xb$ and $\xw$.

Figures~\ref{fig:fptd095} and \ref{fig:fptd07} present survival probabilities for a system in which there is a competition between long waiting times and long jumps. The survival probabilities depicted in Figs.~\ref{fig:fptd095} and \ref{fig:fptd07} have been rescaled in such a way that medians of first passage times densities are located at the same point, see \cite{dybiec2010}. The insets show the behavior of the survival probability at small values of first passage times. Furthermore, the solid lines present the dependence of the survival probability for the escape from the finite interval of width $L=1$, see \cite{dybiec2010}, whose asymptotics is of the same type as the asymptotic dependence of the survival probability in the generalized Kramers problem. In the case of both the anomalous diffusion on finite intervals and the Kramers problem the asymptotic power-law behavior is determined by the waiting time exponent $\nu$. Figs.~\ref{fig:fptd095} and \ref{fig:fptd07} confirm that the asymptotic dependence of survival probabilities is the same regardless of the exact value of the stability index $\alpha$, see Eq.~(\ref{eq:sasymptotic}). The survival probability from the finite intervals and survival probabilities for the Kramers problem differ at small values of first passage times, see insets in Figs.~\ref{fig:fptd095} and \ref{fig:fptd07}. In the limit of the waiting time exponent $\nu=1$, the studied process becomes a Markovian L\'evy flight and the survival probability attains its exponential asymptotic dependence, see Eq.~(\ref{eq:sasymptotic1}) and \cite{eliazar2004,chechkin2005,imkeller2006b,dybiec2007,Chechkin2007}.

The asymptotic power-law decay of the survival probability $S(t)$, see Eq.~(\ref{eq:sasymptotic}) and Figs.~\ref{fig:fptd095} and \ref{fig:fptd07} is the direct consequence of the properties of the Mittag-Leffler function which describes decay of single modes of the fractional Fokker-Planck equation (\ref{eq:ffpe}). The power-law decay observed in non-Markovian situations ($\nu<1$) can be contrasted with Markovian L\'evy flights ($\nu=1$) associated with the exponential asymptotics of the survival probability.
Moreover, non-Markovian properties of the system studied are not only demonstrated by a slow decay of the survival probability but also by a long memory of initial conditions \cite{sokolov2002,dybiec2009h,dybiec2010}. Both long memory signatures mentioned originate in temporal properties of the fractional Fokker-Planck equation. In other words, non-Markovian L\'evy flights (anomalously long jumps coexisting with long waiting times) do not diminish the memory effects of the process studied. The presence of long jumps can be observed in the tail asymptotics of $p(x,t)$, see \cite{dybiec2009h,dybiec2010b}. In such a situation the waiting time exponent $\nu$ determines the ratio of convergence to the asymptotic dependence or to stationary states \cite{dybiec2010b}. Finally, the slowly decaying memory (controlled by the exponent $\nu$) is responsible for a slow asymptotic decay of the survival probability (characterized by the exponent $\nu<1$) and the divergence of the mean first passage time (as in the case of escape from finite intervals \cite{yuste2004,dybiec2010}).

Figures~\ref{fig:his095res} and \ref{fig:his07res} present first passage time densities for a small value of first passage times for values of the waiting time exponent $\nu=0.95$ (Fig.~\ref{fig:his095res}) and $\nu=0.7$ (Fig.~\ref{fig:his07res}). First passage time densities are rescaled in the analogous manner as survival probabilities. The rescaled densities, for both the potential well to the barrier top ($\fb$) and the potential well to the potential well ($\fw$) setups, are in the perfect agreement, i.e. curves corresponding to various values of the stability index $\alpha$ are not distinguishable. For small and moderate values of the waiting time exponent, the short time behavior of the first passage time densities agrees with the short time asymptotics derived from the short time form of the Mittag-Leffler function (solid lines), see Eq.~(\ref{eq:mlfstretched}). However, the level of agreement is better for the potential well to the barrier top setup than for the potential well to the potential well configuration. Furthermore, with the decrease of the waiting time exponent the level of agreement increases, see Figs.~\ref{fig:his095res} and \ref{fig:his07res}, because a decreasing value of the waiting time exponent increases the probability of long waiting times to occur.

In a general case, the asymptotic of the survival probability is determined by the value of the waiting time exponent $\nu$ and properties of the Mittag-Leffler functions. For large values of the first passage time $t$, survival probabilities behave like $t^{-\nu}$ ($\nu<1$), see also \cite{Metzler2000c}. Consequently, analogously as for the escape from the finite intervals \cite{yuste2004,dybiec2010}, the mean first passage time from the potential well diverges. Due to a slow power-law decay of the first passage time density it cannot be characterized by either the mean value or the variance. Therefore, in order to further elucidate properties of the first passage time density we investigate properties of the median and width of the first passage time distributions. The median is the quantile $q_{0.5}$ of the first passage time density, i.e. it is defined by $\int_0^{q_{0.5}}f(t)dt=0.5$ or equivalent $S(q_{0.5})=0.5$. To characterize the width of the distribution we use the interquantile distance $q_{0.9}-q_{0.1}$, i.e. the width of the interval containing 80\% of first passage times.

Figure~\ref{fig:medianaalpha} presents the location of the median as a function of the stability index $\alpha$ for the potential well to the barrier top ($q^{\mathrm{w-b}}_{0.5}$ -- left panel) and the potential well to the potential well ($q^{\mathrm{w-w}}_{0.5}$ -- right panel) setups. Fig.~\ref{fig:widthalpha} presents the width of the first passage time density as a function of the stability index $\alpha$ for the potential well to the barrier top ($q^{\mathrm{w-b}}_{0.9}-q^{\mathrm{w-b}}_{0.1}$ -- left panel) and the potential well to the potential well ($q^{\mathrm{w-w}}_{0.9}-q^{\mathrm{w-w}}_{0.1}$ -- right panel) setups. Analogously, Figs.~\ref{fig:mediananu} and \ref{fig:widthnu} present the location of the median and the distribution width as a function of the waiting time exponent~$\nu$.

Both the location of the median ($q_{0.5}$) and the distribution width ($q_{0.9}-q_{0.1}$) are non-monotonic functions of the stability index $\alpha$ with the minima located at $\alpha \approx 1.5$, see Figs.~\ref{fig:medianaalpha} and \ref{fig:widthalpha}.  This effect might be considered as the analogy to the fact that in the presence of (Markovian) L\'evy flights ($\nu=1$) the mean first passage time is a non-monotonic function of the stability index $\alpha$ \cite{dybiec2007}. Due to long waiting times the studied process is characterized by the infinite mean first passage time, however, the median ($q_{0.5}$) as a function of the stability index $\alpha$ behaves qualitatively in the same way as the mean first passage time for $\nu=1$ \cite{dybiec2007}. For the potential well to the barrier top setup,  the minima of the median and the distribution width are located at slightly larger values of the stability index $\alpha$ than for the potential well to the (other) potential well setup, see Figs.~\ref{fig:medianaalpha} and \ref{fig:widthalpha}. Finally, in the limit of a small value of the stability index $\alpha$ (with the fixed value of the exponent $\nu$) both setups become indistinguishable.

The dependence of the location of the median ($q_{0.5}$) and the distribution width ($q_{0.9}-q_{0.1}$) as the function of the waiting time exponent $\nu$ is more regular, see Figs.~\ref{fig:mediananu} and \ref{fig:widthnu}. More precisely, locations of the median exhibit some weakly non-monotonic behavior, see bottom curves in Fig.~\ref{fig:mediananu}, while the distribution width decreases with the increase of the exponent $\nu$, see Fig.~\ref{fig:widthnu}. The dependence of the width of the first passage time distribution as a function of the waiting time exponent is typical, i.e. the width of the distribution decreases with the increase of the exponent $\nu$. This monotonic dependence can be explained by the fact that with increasing $\nu$, long waiting times for the next step become less probable, which in turn leads to a decrease in the fraction of extreme events and decrease of the distribution width.

Despite the fact that for $\nu<1$ the mean first passage times ($\MFPTb$ and $\MFPTw$) diverge, it is still possible to find an optimal value of the stability index $\alpha$. The optimal value of the stability index $\alpha$ leads to the minimal value of the median (see Fig.~\ref{fig:medianaalpha}) or the minimal width of the first passage time distributions (see Figs.~\ref{fig:widthalpha}). This non-monotonic dependence of median locations or distributions' widths can be explained by the non-equilibrium character of L\'evy noises \cite{brockmann2002,delcastillonegrete2008,*dybiec2008d,*dybiec2008e}. On the one hand, following the line of reasoning sketched in \cite{brockmann2002}, fluctuations in temperature  in the Langevin equation driven by the white Gaussian noise ($\xi(t)$), i.e.
$ \frac{dx}{dt}=\sqrt{2 T(t)}\xi(t)$,
result in the occurrence of L\'evy noises. In other words, if fluctuations in temperature $T(t)$ are ruled by a one sided L\'evy process $L_{\alpha/2,1}(t)$, it is possible to derive the fractional Fokker-Planck equation (\ref{eq:ffpe}) with the spatial fractional derivative using the Langevin equation  with the fluctuating temperature \cite{brockmann2002}. On the other hand, fluctuations in temperature alter the height of the potential barrier, because within equilibrium statistical mechanics the height of the potential barrier can be related to temperature. Therefore, the system studied resembles a classical model of the resonant activation \cite{doering1992} and the mean first passage time (for $\nu=1$), as well as the median and the width of the first passage time distribution are non-monotonic function of the stability index $\alpha$. Nevertheless, this analogy is very loose. On the contrary to the Gaussian white noise, due to the non-equilibrium character of white L\'evy noises (with $\alpha < 2$) the distribution width is not related to temperature.

%
%
\section{Conclusions}

Using the recently developed scheme of the subordination \cite{magdziarz2007b,magdziarz2007}, we have studied the Kramers problem in the case of non-Markovian L\'evy flights, i.e. long waiting times coexisting with long jumps distributed according to $\alpha$-stable densities. The model considered exhibits a behavior similar to escape from the positive semi-line \cite{Balakrishnan1985,*rangarajan2000,*barkai2001,koren2007c,dybiec2009d} and finite intervals \cite{dybiec2010} when the asymptotic behavior of the first passage time distribution is not sensitive to the details of the jump length distribution but is determined by the waiting time exponent~$\nu$.

Using analytical and numerical arguments, we have demonstrated that the asymptotic power-law behavior of the survival probability is a general feature of the non-Markovian L\'evy flights which are characterized by anomalously long waiting times distributed according to a probability density with a power-law asymptotics and jump lengths distributed according to the $\alpha$-stable density. Analytical arguments are provided by properties of the Riemann-Liouville fractional time derivative which leads to the occurrence of the Mittag-Leffler decay pattern, while the numerical arguments are based on the Monte Carlo simulations of trajectories of the stochastic process otherwise described by the fractional Fokker-Planck equation. The numerical results constructed by the subordination method (which provides a stochastic representation of solutions of the fractional Fokker-Planck equation) corroborated analytical predictions. Furthermore, by a proper rescaling of first passage time densities, we have demonstrated their universal character not only for large values of first passage times but also for short times.

For the fractional Kramers problem, the mean first passage time characterizing the escape of the particle over the potential barrier is infinite. The survival probability is asymptotically characterized by the exponent which is equal to the value of the exponent characterizing asymptotic power-law decay of the waiting time distributions. The power-law decay of the survival probability is not fast enough to ensure a finite value of the mean first passage time. The presence of long jumps (coexisting with long waiting times) does not change the asymptotic behavior of the survival probability. Consequently, despite the presence of long jumps, the mean first passage time for the fractional Kramers problem remains infinite.

For the value of the waiting time exponent $\nu<1$, the mean first passage time cannot be used to characterize the fractional Kramers problem. The escape of the particle from the potential well can be quantified by properties of the first passage time density. The examination of the first passage time density shows that both the median and the width (defined as the interquantile distance) of the first passage time density depend in a non-monotonic way on the value of the stability index $\alpha$. There exists such a value of the exponent $\alpha$ characterizing the jump length distribution for which the distribution width is minimal and also the median is minimal. This non-monotonic behavior of the median and the distribution width is the consequence of a non-equilibrium character of the $\alpha$-stable L\'evy type noises.


%
%

\end{document}